\documentclass[prd]{revtex4}
\usepackage{amsmath,amssymb}
\usepackage{xcolor}

\numberwithin{equation}{section}
\begin{document}
\title{Magnetic corrections to the fermionic Casimir effect with Lorentz symmetry violation and a compactified extra dimension}
\author{Andrea Erdas}
\email{aerdas@loyola.edu}
\affiliation{Department of Physics, Loyola University Maryland, 4501 North Charles Street,
Baltimore, Maryland 21210, USA}
%\date{November 8, 2026}
\begin {abstract} 
In this work I study the effect of a magnetic field on the Casimir effect due to a massive, charged fermion field that violates Lorentz invariance in an aether-like manner while maintaining CPT, in a space with one toroidally compactified extra dimension. I take the fermion field to obey MIT bag boundary conditions on two identical square parallel plates. I do this investigation using the zeta function technique that enables me to calculate the Casimir energy and pressure for a massless fermion, and one with a light mass, in the presence of a  constant magnetic field perpendicular to the plates. I examine the cases of timelike Lorentz violation, of spacelike Lorentz violation in the direction of the magnetic field, of spacelike Lorentz violation in the direction perpendicular to the magnetic field, and of spacelike Lorentz violation in the direction of the compactified extra dimension. In all scenarios, I find simple and accurate analytic expressions of the magnetic field dependent Casimir energy and pressure.
%%%%%%%%%%%%%%%%%%%%%%%%%%%%%%%%%%%%%%%%%%%%%%%%%%%%%%%%%%%%%%%%
\end {abstract}
%\pacs{03.70.+k, 11.10.-z, 11.30.Cp, 12.20.Ds.}
\maketitle
%%%%%%%%%%%%%%%%%%%%%%%%%%%%%%%%%%%%%%%%%%%%%%%%%%%%%%%%%%%%%
\section{ Introduction}
\label{1}
The Casimir effect is a theoretical prediction of Hendrik Casimir \cite{Casimir:1948dh}, made 78 years ago. 
In its original form, he showed that the effect of quantum field theory is to produce an attraction between two conducting plates facing each other in a vacuum. The first weak experimental confirmation of it \cite{Sparnaay:1958wg} came ten years after Casimir's work and was loosely consistent with his theoretical prediction. 
Many experimental tests have been done \cite{Bordag:2001qi,Bordag:2009zz,Bressi:2002fr} in the decades that followed Casimir's  paper, producing more accurate and successful confirmation of his prediction. Casimir's work focused only on quantum fluctuations of the electromagnetic field in vacuum and discovered they cause an attractive force between the plates.
Casimir forces are produced also by quantum fluctuations of other fields and depend strongly on the boundary conditions at the plates and on the shape of the plates \cite{Boyer:1974,Boyer:1968uf}. Boundary conditions such as 
Dirichlet, Neumann or mixed are suitable for scalar and electromagnetic fields and have been used extensively when studying the Casimir effect due to fields of that kind, but are not suitable for fermion fields \cite{Ambjorn:1981xv}. Bag boundary conditions, initially proposed as a solution to confinement, are suitable and used for fermion fields \cite{Chodos:1974je,Johnson:1975zp} in the context of the Casimir effect.

Lorentz symmetry violation has been the subject of vigorous theoretical investigations in the last twenty years, with the proposal of several models that cause space-time anisotropy  \cite{Horava:2009uw,Ferrari:2010dj,Ulion:2015kjx}, within the context of proposals of a theory of quantum gravity, or of string theory. These models generally break Lorentz symmetry at the Planck scale, but repercussions of the  anisotropy of space-time can be observed at much lower energy through, for example, the Casimir effect, in the case of Lorentz symmetry breaking scalar fields \cite{Cruz:2017kfo,Cruz:2018bqt} and fermion fields \cite{daSilva:2019iwn} within the context of the Horava-Lifshitz model. In this paper, I focus on a fermion field model with a CPT even, aether-like Lorentz symmetry violation, such as that used in Refs. \cite{Gomes:2009ch,Cruz:2018thz,Rohim:2025ial,deFarias:2023xjf}.

The possibility of compactified extra dimensions has been suggested first by Kaluza and Klein over a century ago, as an attempt to unify gravity and electromagnetism \cite{Kaluza:1921tu,Klein:1926tv}. This revolutionary idea has been resurrected  in 1998 by the famous papers of Arkani-Hamed, Dimopoulos and Dvali \cite{Arkani-Hamed:1998jmv,Antoniadis:1998ig,Arkani-Hamed:1998sfv}, where they propose large compact extra dimensions within a new framework, embedded in string theory, as a possible solution to the hierarchy problem. The size of the compactified extra dimensions that these authors propose ranges between one millimeter and one TeV$^{-1}$. One year after the publication of these papers, Randall and Sundrum \cite{Randall:1999ee} obtained another solution to the hierarchy problem that involves a five-dimensional AdS spacetime with warped geometry. Gravity is modified by these proposed scenarios, producing deviations from Newton's universal law of gravity at the sub-millimeter scale. Gravity at the sub-millimeter scale has been tested by many experiments in the last twenty years, see Refs. \cite{Adelberger:2009zz,Murata:2014nra} for  reviews of representative experiments and constraints, and deviations from Newton's gravity are placed at or below $40\mu$m. There are also cosmological and astrophysical constraints on the size and number of compactified extra dimensions, as well as constraints from LHC. Furthermore, a strong link between compactified extra dimensions and Lorentz violation is provided by Ref. \cite{Rizzo:2005um}, where the author shows that in theories with Kaluza-Klein compactified extra dimensions sources of Lorentz violation must exist that are associated with the physics that produced the compactification process at high energy. See also  \cite{Colladay:1998fq} for a specific model of this, and \cite{Carroll:2008pk} for a connection between Lorentz violating aether models and compactified extra dimensions. In this paper I work with one toroidally compactified extra dimension, following the compactification process introduced in Ref. \cite{Bellucci:2009hh}.

Recently, authors have studied magnetic correction to the Casimir effect caused by charged scalar fields \cite{Cougo-Pinto:1998jun,Cougo-Pinto:1998fpo,Erdas:2013jga,Erdas:2013dha} and by Majorana fermion fields \cite{Erdas:2010mz} in the standard four-dimensional space-time without Lorentz violation. Magnetic corrections to the Casimir effect caused by a charged scalar field that breaks Lorentz invariance \cite{Erdas:2020ilo,Erdas:2021xvv,Erdas:2025gbv,Droguett:2025frq} and by a fermion field that breaks Lorentz symmetry \cite{Erdas:2023wzy} have also been studied, all in four-dimensional space-time. An interesting work was published last year \cite{Rohim:2025ial}, where the authors investigate the Casimir effect of a Lorentz-violating fermion field in a space with one compactified extra dimension. However, a study of the magnetic effects on the Casimir effect caused by a charged fermion field that breaks Lorentz invariance in a space with one compactified extra dimension has not been done. In this work I will investigate how a uniform magnetic field affects the Casimir effect due to a charged fermion field  with light mass, in a space with one compactified extra dimension.
This field violates Lorentz symmetry and satisfies MIT bag boundary conditions on two square parallel plates of the same size and facing each other, while the uniform magnetic field is perpendicular to the plates. I will examine the case of a massless fermion field and one with light mass.

In Sec. \ref{2} of this work I introduce the model of a massive fermion field that breaks Lorentz invariance in a CPT-even aether-like manner and is in space with a toroidally compactified extra dimension. This field satisfies bag boundary conditions on the two plates. I first obtain the field vacuum energy without a magnetic field and then in the presence of the magnetic field for all possibilities of direction of the Lorentz violation. 
In Sec. \ref{3}, I use the zeta function technique to calculate the Casimir energy of this system in the presence of the external magnetic field for representative orientations of Lorentz violation relative to the magnetic field, the plate normal, and the compactified dimension. I examine the cases of a fermion field with vanishing mass, and light mass. In Sec. \ref{4}, I obtain the Casimir pressure on the plates for representative orientations of Lorentz violation, as stated above, and for vanishing fermion mass and light mass. I present a summary and discussion of my results in Sec. \ref{5}. 
%%%%%%%%%%%%%%%%%%%%%%%%%%%%%%%%%%%%%%%%%%%%%%%%%%%%%%%%%%%%%%%%%%%%%
\section{The model}
\label{2}

In this work I study the Casimir effect due to a fermion field $\psi$ of charge $e>0$ and mass $m$ that violates Lorentz symmetry in a CPT-even aether-like manner according to the model proposed in \cite{Gomes:2009ch,Cruz:2018thz,Carroll:2008pk}. In addition to Lorentz violation, I consider a five-dimensional spacetime where the fifth dimension is toroidally compactified,  $R^{4}\times S^1$, where $R^{4}$ is the standard four-dimensional Minkowski space time and the extra dimension $S^1$ is compactified onto a circle, as in Ref. \cite{Rohim:2025ial,Bellucci:2009hh}. I use $\hbar=c=1$, and  $g_{\mu\nu}\equiv {\rm diag.}(+1,-1,-1,-1)$ as the  metric tensor for Minkowski spacetime throughout this paper. 

The Lagrangian for the Lorentz violating fermion field $\psi$ in the space with one compactified extra dimension described above and, for now, without a magnetic field, is shown in Ref. \cite{Rohim:2025ial}
\begin{equation}{\cal L}={\bar \psi}(i\gamma^a\partial_a-m +i\lambda u^a\gamma_a u^b \partial_b)\psi
\label{L}
\end{equation}
where $a=0,1,2,3,4$; $\gamma^a=(\gamma^\mu, \Gamma^5)$ are the Dirac gamma matrices and $\Gamma^5=\gamma^0\gamma^1\gamma^2\gamma^3$ is the Clifford algebra generator for the compactified fifth dimension, $u^a$ is a unit vector pointing in the direction of the Lorentz violation, and $\lambda\ll1$, is a dimensionless parameter that quantifies the amount of Lorentz violation. Notice that, with the choice of $\gamma^4=\Gamma^5$ stated above, the anticommutation relation for the $\gamma^a$ reads $\{\gamma^a,\gamma^b\} = 2g^{ab}$, with $g_{ab}\equiv {\rm diag.}(+1,-1,-1,-1,-1)$.
The  modified Dirac equation for $\psi$ is also obtained in Ref. \cite{Rohim:2025ial}
\begin{equation}(i\gamma^a\partial_a-m +i\lambda u^a\gamma_a u^b \partial_b)\psi=0,
\label{DL}
\end{equation}
and the compactified extra dimension $x^4$ is such that
\begin{equation}\psi(x^\mu,x^4+q)=e^{2i\pi\beta}\psi(x^\mu,x^4)
\label{x4}\end{equation}
where $q$ is the size of the compactified extra dimension, and $\beta$ the compactification parameter, $0\le\beta<1$.

Next, I consider two square parallel plates of area $L^2$ perpendicular to the $x^3$-axis, one located at $x^3=0$ and the other at $x^3=a$, with $a\ll L$, and impose MIT bag boundary conditions for the fermion field on the plates
\begin{equation}in_a\gamma^a\psi=\psi
\label{bag}
\end{equation}
where $n_a$ is the unit vector perpendicular to the plate. These boundary conditions cause the momentum component of $\psi$ in the $x^3$-direction, $k^3$, to be discretized, as we'll see below.

The general expression of the vacuum energy $E_{V\!\! AC} = \langle 0 \vert \hat{H}\vert 0 \rangle$ for the Lorentz-violating quantum field plus plates configuration in the five-dimensional space described above is obtained in Ref. \cite{Rohim:2025ial} for various directions of $u^a$, the unit vector that points in the direction of the Lorentz violation. Below, I list the vacuum energy for the representative directions of $u^a$.
\begin{enumerate}
\item
When the Lorentz violation is timelike, $u^a=(1,0,0,0,0)$, 
the Lagrangian of Eq. (\ref{L}) becomes
\begin{equation}{\cal L}={\bar \psi}\left[i(1+\lambda)\gamma^0\partial_0+i\gamma^j\partial_j-m \right]\psi,
\label{Lt}
\end{equation}
and the modified Dirac equation of Eq. (\ref{DL}) becomes 
\begin{equation}\left[i(1+\lambda)\gamma^0\partial_0+i\gamma^j\partial_j-m \right]\psi=0,
\label{DLt}
\end{equation}
where the summation index $j$ goes from one to four. The momentum density conjugate to $\psi$ is therefore
\begin{equation}\pi={\partial {\cal L}\over\partial(\partial_0\psi)}=i(1+\lambda)\psi^\dagger,
\label{Pi}
\end{equation}
and the Hamiltonian is
\begin{equation}H=\int d^4x\left[\pi(\partial_0\psi)-{\cal L}\right]=i(1+\lambda)\int d^4x\,\psi^\dagger(\partial_0\psi).
\label{H}
\end{equation}
Now I quantize the fermion field in open four-dimensional space to obtain the Hamiltonian operator and then, from it, the vacuum energy. Once I obtain the vacuum energy, it will be easy to modify it to the case of a finite-sized three-dimensional region with a compactified fourth-dimension. Adopting the standard procedure found in all quantum field theory textbooks \cite{Peskin:1995ev,Mandl:2010}, I start by writing the fermion field operator and its hermitian conjugate following closely the notation of Ref. \cite{Peskin:1995ev}
\begin{equation}
\hat{\psi}(x)=\int{d^4k\over(2\pi)^4}{1\over\sqrt{2\omega}}\sum_s\left[a^s({\bf k})u^s(k)e^{-i k\cdot x}+b^{s\dagger}({\bf k})v^s(k)e^{i k\cdot x}
\right],
\label{psihat}
\end{equation}
\begin{equation}
\hat{\psi}^\dagger(x)=\int{d^4k\over(2\pi)^4}{1\over\sqrt{2\omega}}\sum_s\left[a^{s\dagger}({\bf k})u^{s\dagger}(k)e^{i k\cdot x}+b^{s}({\bf k})v^{s\dagger}(k)e^{-i k\cdot x}
\right],
\label{psihatdagger}
\end{equation}
where $k^a=(\omega,{\bf k})$ and
\begin{equation}
\omega=(1+\lambda)^{-1}\sqrt{m^2+k_1^2+k_2^2+k_3^2+k_4^2}.
\label{omega}
\end{equation}
The spin sum index is $s$, $u^s(k)$ and $v^s(k)$ are the positive and negative-frequency spinors, and $a^s({\bf k})$, $a^{s\dagger}({\bf k})$, $b^s({\bf k})$, $b^{s\dagger}({\bf k})$ are annihilation and creation operators for fermions and anti-fermions respectively, that obey the following equal-time anticommutation rules
\begin{equation}
\left\{a^r({\bf k}),a^{s\dagger}({\bf k'}) \right\}=\left\{b^r({\bf k}),b^{s\dagger}({\bf k'}) \right\}=(2\pi)^4\delta^{(4)}({\bf k}-{\bf k'})\delta^{rs},
\label{anticom}
\end{equation}
with all other anticommutators equal to zero. These anticommutation rules yield the following equal-time anticommutation relations for the field operator $\hat \psi$ and its hermitian conjugate 
\begin{equation}
\left\{{\hat \psi}_a({\bf x}),{\hat \psi}^\dagger_b({\bf y}) \right\}=\delta^{(4)}({\bf x}-{\bf y})\delta_{ab},
\label{psianticom}
\end{equation}
\begin{equation}
\left\{{\hat \psi}_a({\bf x}),{\hat \psi}_b({\bf y}) \right\}=\left\{{\hat \psi}^\dagger_a({\bf x}),{\hat \psi}^\dagger_b({\bf y}) \right\}=0,
\label{psianticom2}
\end{equation}
where $a$ and $b$ are spinor indices. Since $\hat \pi$ is related to ${\hat \psi}^\dagger$ by Eq. (\ref{Pi}), its anticommutator with ${\hat \psi}$ is
\begin{equation}
\left\{{\hat \psi}_a({\bf x}),{\hat \pi}_b({\bf y}) \right\}=i(1+\lambda)\delta^{(4)}({\bf x}-{\bf y})\delta_{ab}.
\label{cananticom}
\end{equation}
Next, I insert the expressions of ${\hat \psi}$ and ${\hat \psi}^\dagger$ from Eqs. (\ref{psihat}) and (\ref{psihatdagger}) into Eq. (\ref{H}) and, after some tedious algebra whose details are shown in the Ref.  \cite{Peskin:1995ev} graduate textbook, I obtain the Hamiltonian operator
\begin{equation}
{\hat H}=(1+\lambda)\int{d^4k\over(2\pi)^4}\omega\sum_s\left[a^{s\dagger}({\bf k})a^s({\bf k})-b^s({\bf k})b^{s\dagger}({\bf k})\right].
\label{Hop}
\end{equation}
I indicate with $|0>$ the ground state of $\hat H$, then
\begin{equation}
<0|a^{s\dagger}({\bf k})a^s({\bf k})|0>=0 \;\;\;;\;\;\;<0|b^s({\bf k})b^{s\dagger}({\bf k})|0>=1,
\label{Hop2}
\end{equation}
and find the vacuum energy $E_{V\!\!AC}=<0|{\hat H}|0>$,
\begin{equation}
<0|{\hat H}|0>=-2(1+\lambda)\int{d^4k\over(2\pi)^4}\omega,
\label{Hop3}
\end{equation}
where the factor of two is due to the sum over the two spin states. Notice how the factor of $(1+\lambda)$ is canceled by the factor of $(1+\lambda)^{-1}$ that is present in Eq. (\ref{omega}) for $\omega$, and therefore the vacuum energy is
\begin{equation}
<0|{\hat H}|0>=-2\int{d^4k\over(2\pi)^4}\sqrt{m^2+k_1^2+k_2^2+k_3^2+k_4^2}.
\label{Hop4}
\end{equation}
Once I include the two Casimir plates, the fermion field must satisfy MIT bag boundary conditions at the plates, producing the following constraint on $k_3$
\begin{equation}
ma\sin (k_3 a)+k_3a\cos(k_3 a) =0,
\label{MIT}
\end{equation}
which is valid for every Kaluza-Klein mode, indicated with $\ell$ in the equation below, and for every Landau level $j$, when I will introduce the magnetic field and its Landau levels in Eq. (\ref{E_pm1}). Eq. (\ref{MIT}) has been obtained by the authors of Ref. \cite{Rohim:2025ial} in their equation (18), who show explicitly that this transcendental equation depends on $m$ and not on $\sqrt{m^2+k^2_\ell}$.
I call $\alpha_n$ the $n$-th root of the above transcendental equation and introduce the dimensionless quantity $k_n=a\alpha_n$ and, clearly, $n\ge 1$. Once I impose the quasi-periodic boundary conditions of Eq. (\ref{x4}), $k_4$ becomes discretized and takes the following values
 \begin{equation}
k_\ell={2\pi\over q}(\ell+\beta),
\label{kl}
\end{equation}
with $\ell=0, \pm 1, \pm 2, \cdots$ .
Therefore, I use Eqs. (\ref{Hop4}) to write the vacuum energy in the presence of the two plates and with the fourth spatial dimension being toroidally compactified
\begin{equation}E_{V\!\!AC} = -C{L^2\over (2\pi)^2}\int dk_1 dk_2\sum_{\ell=-\infty}^\infty\sum_{n=1}^\infty\left[m^2+k_1^2+k_2^2+
\left({k_n\over a}\right)^2+k^2_\ell
\right]^{1/2},
\label{Evac1}
\end{equation}
where $C=2$ is the spin degeneracy factor,  $k_n=a\alpha_n$, and $k_\ell$ is defined in Eq. (\ref{kl}). Notice that $k_3$ is discretized due to the presence of the Casimir plates and $k_4$ is discretized because this dimension is compactified. The vacuum energy in this form is in full agreement with Ref. \cite{Rohim:2025ial} and also with Ref. \cite{Cruz:2018thz} that deals with the same model of Lorentz violating fermion field but in three spatial dimensions, without the compactified extra dimension.
\item
When the Lorentz violation is spacelike and in the $x^3$-direction, $u^a=(0,0,0,1,0)$,
\begin{equation}E_{V\!\!AC} = -C{L^2\over (2\pi)^2}\int dk_1 dk_2\sum_{\ell=-\infty}^\infty\sum_{n=1}^\infty\left[m^2+k_1^2+k_2^2+
(1-\lambda)^2\left({k_n\over a}\right)^2+k^2_\ell
\right]^{1/2},
\label{Evac2}
\end{equation}
the equation above is derived with a method very similar to the one I show above for the timelike case, and I will not show again that derivation here. What changes in this case is the transcendental equation of Eq. (\ref{MIT})
that discretizes $k_3$ and becomes
\begin{equation}
ma\sin (k_3 a)+(1-\lambda)k_3a\cos(k_3 a) =0,
\label{MIT2}
\end{equation}
this transcendental equation, similarly to Eq. (\ref{MIT}), is valid for every Kaluza-Klein mode $\ell$ and for every Landau level $j$, it depends on $m$ and not on $\sqrt{m^2+k^2_\ell}$, and it is obtained in equation (36) of Ref. \cite{Rohim:2025ial}.
\item
When the Lorentz violation is spacelike and in the $x^1-x^2$ direction, $u^a=(0,{1\over\sqrt{2}},{1\over\sqrt{2}},0,0)$,
\begin{eqnarray}E_{V\!\!AC} &=& -C{L^2\over (2\pi)^2}\int dk_1 dk_2
\nonumber \\
&\times&\sum_{\ell=-\infty}^\infty\sum_{n=1}^\infty\left[m^2+\left(1-{\lambda\over 2}\right)^2(k_1^2+k_2^2)+
\left({k_n\over a}\right)^2+k^2_\ell
\right]^{1/2}.
\label{Evac3}
\end{eqnarray}
\item
When the Lorentz violation is spacelike and in the direction of the compactified extra dimension, $u^a=(0,0,0,0,1)$,
\begin{equation}E_{V\!\!AC} = -C{L^2\over (2\pi)^2}\int dk_1 dk_2\sum_{\ell=-\infty}^\infty\sum_{n=1}^\infty\left[m^2+k_1^2+k_2^2+
\left({k_n\over a}\right)^2+(1-\lambda)^2k^2_\ell
\right]^{1/2}.
\label{Evac4}
\end{equation}
\end{enumerate}

The novelty of this work is the presence of a constant magnetic field  in the region where the Casimir plates are located. The magnetic field has magnitude $B$ and is perpendicular to the plates, thus pointing in the $x^3$-direction. The Lagrangian of Eq. (\ref{L}), in the presence of a constant magnetic field pointing in the $x^3$-direction, needs to be changed. The derivative $\partial_a$ needs to be replaced by the electromagnetically covariant derivative $D_a=\partial_a+ieA_a$, where the electromagnetic vector potential is
\begin{equation}A^a=(0,0,Bx^1,0,0).
\label{A_a}
\end{equation}
Notice that, to preserve manifest gauge covariance of the Lagrangian, this replacement must occur in the ordinary Dirac term and the Lorentz-violating term of Eq. (\ref{L}). Following the procedure outlined, for example, in Refs. \cite{Erdas:2013jga,Erdas:2010mz}, I find that
the presence of the magnetic field removes the spin degeneracy and, instead of real-valued $k_1$ and $k_2$, discrete Landau levels appear. The vacuum energy therefore becomes
\begin{equation}
E_{V\!\!AC}=E_++E_-, 
\label{E_pm}
\end{equation}
where $E_+$ is the contribution of the fermion field spin up component to the vacuum energy, and $E_-$ the contribution of the spin down component. I find:
\begin{enumerate}
\item
when the Lorentz violation is timelike
\begin{equation}
E_\pm=-{L^2 eB \over 2 \pi}\sum_{\ell=-\infty}^\infty\sum_{n=1}^\infty\sum_{j=0}^\infty \left[m^2+(2j+1\pm 1)eB+\left({k_n\over a}\right)^2+k^2_\ell \right]^{1/2},
\label{E_pm1}
\end{equation}
where $j = 0,1,2,\cdots$ labels the Landau levels, $(2j+1)eB$ is the energy of the $j$-th Landau level, $\pm eB$ is the spin contribution, ${L^2 eB \over 2 \pi}$ takes into account the degeneracy of the Landau levels and, with the definitions of $e$ and $B$ given above, $eB>0$.
\item
When the Lorentz violation is spacelike in the $x^3$-direction
\begin{equation}
E_\pm=-{L^2 eB \over 2 \pi}\sum_{\ell=-\infty}^\infty\sum_{n=1}^\infty\sum_{j=0}^\infty \left[m^2+(2j+1\pm 1)eB+(1-\lambda)^2\left({k_n\over a}\right)^2+k^2_\ell \right]^{1/2},
\label{E_pm2}
\end{equation}
\item
when the Lorentz violation is spacelike in the $x^1-x^2$ direction
\begin{eqnarray}
E_\pm&=&-{L^2 eB \over 2 \pi}\sum_{\ell=-\infty}^\infty\sum_{n=1}^\infty\sum_{j=0}^\infty \left[m^2+\left(1-{\lambda\over 2}\right)^2 \right.
\nonumber \\ 
&\times&\left.
(2j+1\pm 1)eB+\left({k_n\over a}\right)^2+k^2_\ell \right]^{1/2},
\label{E_pm3}
\end{eqnarray}
\item
when the Lorentz violation is spacelike in the direction of the compactified extra dimension
\begin{equation}
E_\pm=-{L^2 eB \over 2 \pi}\sum_{\ell=-\infty}^\infty\sum_{n=1}^\infty\sum_{j=0}^\infty \left[m^2+(2j+1\pm 1)eB+\left({k_n\over a}\right)^2+(1-\lambda)^2k^2_\ell \right]^{1/2}.
\label{E_pm4}
\end{equation}
\end{enumerate}

%%%%%%%%%%%%%%%%%%%%%%%%%%%%%%%%%%%%%%%%%%%%%%%%%%%%%%%%%%%%%%%%%%%%%
\section{Casimir energy}
\label{3}
In this section, I evaluate the Casimir energy of the field $\psi$, using the zeta function technique \cite{Hawking:1976ja}. In subsection \ref{2_1}, I obtain the Casimir energy for the case of timelike Lorentz invariance violation, in subsection \ref{2_2} I use the result of subsection \ref{2_1} to obtain the Casimir energy for the three cases of spacelike Lorentz  violation.
%%%%%%%%%%%%%%%%%%%%%%%%%%%%%%%%%%%%%%%%%%%%%%%%%%%%%%%%%%%%%%%%%%%%%
\subsection{Timelike Lorentz invariance violation}
\label{2_1}
I begin by obtaining $E_{V\!\!AC}$ for the case in which the Lorentz violation is timelike.
I define the parameter $s$ as
\begin{equation}
s=-{{1+\epsilon}\over2},
\label{s}
\end{equation}
then write Eq. (\ref{E_pm1}) as
\begin{equation}
E_\pm=-\lim_{\epsilon \rightarrow 0}{L^2 eB \over 2 \pi}\sum_{\ell=-\infty}^\infty\sum_{n=1}^\infty\sum_{j=0}^\infty \left[m^2+(2j+1\pm 1)eB+\left({k_n\over a}\right)^2+k^2_\ell \right]^{-s},
\label{E_pm5}
\end{equation}
and, using the following identity
\begin{equation}
z^{-s}=
{1\over\Gamma(s)}\int_0^\infty dt t^{s-1}e^{-zt},
\label{z}
\end{equation}
I rewrite $E_\pm$ for the case of timelike Lorentz violation as
\begin{equation}
E_\pm=-\lim_{\epsilon \to 0}{L^2 eB \over 2 \pi\Gamma(s)}\sum_{\ell=-\infty}^\infty\sum_{n=1}^\infty\sum_{j=0}^\infty \int_0^\infty dt \,t^{s-1}e^{-\left[m^2+\left({k_n\over a}\right)^2+k^2_\ell \right]t}  \ e^{-(2j+1\pm 1)eBt},
\label{E_pm6}
\end{equation}
and obtain $E_{V\!\!AC}=E_++E_-$,
\begin{equation}
E_{V\!\!AC}=-\lim_{\epsilon \to 0}{L^2 eB \over 2 \pi\Gamma(s)}\sum_{\ell=-\infty}^\infty\sum_{n=1}^\infty \int_0^\infty dt \,t^{s-1}e^{-\left[m^2+\left({k_n\over a}\right)^2+k^2_\ell \right]t}  \coth(eBt),
\label{E_03}
\end{equation}
where I used the other identity
\begin{equation}
\sum_{j=0}^\infty \left[e^{-2(j+1)z}+ e^{-2j z} \right]=\coth (z).
\label{coth}
\end{equation}
 At this point I need the explicit form of $k_n$, which I obtained in the Appendix for $ma\ll 1$. I begin by examining the case of light mass, $ma\ll1$. I use the expression of the $k_n$'s obtained in the appendix to find 
 \begin{equation}
k_n = (n-{1\over 2})\pi+ma\left[(n-{1\over 2})\pi
\right]^{-1}+{\cal O}(m^2a^2),
\label{kn}
\end{equation}
 and thus
 \begin{equation}
m^2+\left({k_n\over a}\right)^2 \simeq a^{-2} \left[\left(n-{1\over 2}\right)^2{\pi}^2+2ma+{\cal O}(m^2a^2)\right].
\label{knsq}
\end{equation}
I insert this expression into Eq.  (\ref{E_03}) and then
do a Poisson resummation on the $n$-sum
\begin{equation}
\sum_{n=1}^\infty e^{-\sigma(n-{1\over 2})^2}=\sqrt{\pi\over4\sigma}+\sqrt{\pi\over\sigma}\sum_{n=1}^\infty (-1)^ne^{-{\pi^2n^2\over\sigma}},
\label{Poisson_1}
\end{equation}
to find
\begin{equation}
 \sum_{n=1}^\infty e^{-(n-{1\over 2})^2\left({\pi\over a}\right)^2t}={a\over\sqrt{\pi t}}\left[{1\over 2}+\sum_{n=1}^\infty (-1)^ne^{-{n^2a^2\over t}}\right].
\label{Poisson_2}
\end{equation}
I then insert the Poisson resummation of Eq. (\ref{Poisson_2}) into the vacuum energy expression of Eq. (\ref{E_03}) and obtain
\begin{eqnarray}
E_{V\!\!AC}&=&-\lim_{\epsilon \to 0}{L^2 eB a\over 2 \sqrt{\pi^3}\Gamma(s)}\sum_{\ell=-\infty}^\infty \int_0^\infty dt \,t^{s-3/2}e^{-\left( 2{m\over a}+k_\ell^2 \right)t} \nonumber \\
&\times&
 \left[{1\over 2}+\sum_{n=1}^\infty (-1)^ne^{-{n^2a^2\over t}}\right]\coth(eBt),
\label{E_04}
\end{eqnarray}
where the term that contains only the $\ell$-sum and not the $n$-sum is proportional to $L^2a$, without any other dependence on $a$, and is therefore a uniform energy density term. Since, as I state in Sec.  \ref{2}, the magnetic field is present in the region where the plates are located and thus is inside and outside the plates, bulk contribution terms such as this can be subtracted and will not contribute to the Casimir interaction energy after the subtraction. This term can be neglected, and I obtain
\begin{equation}
E_{V\!\!AC}=-\lim_{\epsilon \to 0}{L^2 eB a\over 2 \sqrt{\pi^3}\Gamma(s)}\sum_{\ell=-\infty}^\infty \int_0^\infty dt \,t^{s-3/2}e^{-\left( 2{m\over a}+k_\ell^2 \right)t}  \sum_{n=1}^\infty (-1)^ne^{-{n^2a^2\over t}}\coth(eBt).
\label{E_05}
\end{equation}
In the equation above, I change integration variable from $t$ to $z$, with $t={zna\over\sqrt{2{m\over a}+k^2_\ell}}$, and find
\begin{eqnarray}
E_{V\!\!AC}&=&-\lim_{\epsilon \to 0}{L^2 eB a\over 2 \sqrt{\pi^3}\Gamma(s)}\sum_{\ell=-\infty}^\infty \sum_{n=1}^\infty (-1)^n\left({na\over\sqrt{2{m\over a}}+k^2_\ell}\right)^{s-1/2}
\nonumber \\
&\times&\int_0^\infty dz \,z^{s-3/2}e^{-\left(z+z^{-1} \right)n\sqrt{ 2{m a}+a^2k^2_\ell}} \coth\left({znaeB\over\sqrt{ 2{m\over a}+k^2_\ell}}\right).
\label{E_06}
\end{eqnarray}

The size of the compactified dimension is certainly smaller than the distance between the plates so, in Eq (\ref{E_06}) I take $q\ll a$. This means that $a^2k_\ell^2\gg 1$ for all values of $\ell$. Thus, in Eq (\ref{E_06}), I only retain the term or terms with the lowest value of $k_\ell$. When $\beta\ne 0$, this lowest value is $k_0={2\pi\beta\over q}$, obtained when $\ell=0$. When $\beta =0$ the situation differs a bit, and I'll examine it later. Therefore, for $\beta\ne 0$, I find
\begin{eqnarray}
E_{V\!\!AC}&=&-\lim_{\epsilon \to 0}{L^2 eB a\over 2 \sqrt{\pi^3}\Gamma(s)} \sum_{n=1}^\infty (-1)^n\left({na\over\sqrt{ 2{m\over a}+k^2_0}}\right)^{s-1/2}
\nonumber \\
&\times&\int_0^\infty dz \,z^{s-3/2}e^{-\left(z+z^{-1} \right)n\sqrt{2{m a}+a^2k^2_0}} \coth\left({znaeB\over\sqrt{2{m\over a}+k^2_0}}\right).
\label{E_07}
\end{eqnarray}
Since $\sqrt{ 2{m a}+a^2k^2_0} \gg 1$, because $a/q\gg 1$ and regardless of the value of $am$, only the $n=1$ term in the sum over $n$ contributes significantly, and I obtain
\begin{eqnarray}
E_{V\!\!AC}&=&\lim_{\epsilon \to 0}{L^2 eB a\over 2 \sqrt{\pi^3}\Gamma(s)} \left({a\over\sqrt{2{m\over a}+k^2_0}}\right)^{s-1/2}
\nonumber \\
&\times&\int_0^\infty dz \,z^{s-3/2}e^{-\left(z+z^{-1} \right)\sqrt{ 2{m a}+a^2k^2_0}} \coth\left({zaeB\over\sqrt{ 2{m\over a}+k^2_0}}\right).
\label{E_08}
\end{eqnarray}
I do the $z$-integration using the saddle point method, and find
\begin{eqnarray}
E_{V\!\!AC}&=&\lim_{\epsilon \to 0}{L^2 eB a\over 2 \pi\Gamma(s)} \left({a\over\sqrt{ 2{m\over a}+k^2_0}}\right)^{s-1/2}
e^{-2\sqrt{2{m a}+a^2k^2_0}}\nonumber \\
&\times& \left(2{m a}+a^2k^2_0\right)^{-1/4}\coth\left({aeB\over\sqrt{ 2{m\over a}+k^2_0}}\right),
\label{E_09}
\end{eqnarray}
and, once I take the limit for $\epsilon\rightarrow 0$, obtain
\begin{equation}
E_{V\!\!AC}=-{L^2 eB \over 4\sqrt{\pi^3 a}} \left(2{m\over a}+k^2_0\right)^{1/4}
e^{-2\sqrt{ 2{m a}+a^2k^2_0}} \coth\left({aeB\over\sqrt{ 2{m\over a}+k^2_0}}\right).
\label{E_10}
\end{equation}
Notice that $E_{V\!\!AC}$ is exponentially suppressed: the larger the ratio $a/q$, the higher the exponential suppression. 
I take $m=0$ and immediately obtain the massless limit of $E_{V\!\!AC}$
\begin{equation}
E_{V\!\!AC}=-{L^2 eB \over 4\sqrt{\pi^3 a}} \left(k_0\right)^{1/2}
e^{-2ak_0} \coth\left({aeB\over{k_0}}\right).
\label{E_10a}
\end{equation}

Now I investigate the case of $\beta =0$.  I start by examining the low mass case. Since $k_0=0$ when $\beta=0$, I retain three terms in the sum over $\ell$ of Eq. (\ref{E_06}): terms with $\ell =-1,0,+1$. If I did not do this and only considered the leading order term with $\ell=0$, I would not observe the effect of the presence of an extra dimension. Thus, I write
\begin{equation}
E_{V\!\!AC}=E_{V\!\!AC}^0+E_{V\!\!AC}^1,
\label{E_11}
\end{equation}
where the leading order term is
\begin{eqnarray}
E_{V\!\!AC}^0&=&-\lim_{\epsilon \to 0}{L^2 eB a\over 2 \sqrt{\pi^3}\Gamma(s)} \sum_{n=1}^\infty (-1)^n\left({na\over \sqrt{2m\over a} }\right)^{s-1/2}
\int_0^\infty dz \,z^{s-3/2}\nonumber \\
&\times&e^{-\left(z+z^{-1} \right)n \sqrt{2m a} } \coth\left({znaeB\over  \sqrt{2m\over a}}\right),
\label{E_12}
\end{eqnarray}
and it does not depend on the extra dimension, while the next to leading order term is
\begin{eqnarray}
E_{V\!\!AC}^1&=&-\lim_{\epsilon \to 0}{L^2 eB a\over  \sqrt{\pi^3}\Gamma(s)} \sum_{n=1}^\infty (-1)^n\left({na\over\sqrt{ {2m\over a} +k^2_1}}\right)^{s-1/2}
\nonumber \\
&\times&\int_0^\infty dz \,z^{s-3/2}e^{-\left(z+z^{-1} \right)n\sqrt{ {2m a}} +a^2k^2_1} \coth\left({znaeB\over\sqrt{ {2m\over a} +k^2_1}}\right),
\label{E_13}
\end{eqnarray}
it has an explicit dependence on the extra dimension, and I use $k_1={2\pi\over q}$.
The leading order term,$E_{V\!\!AC}^0$, independent of the extra dimension, can be found using the methods shown in Ref. \cite{Erdas:2010mz}. In the limit of light fermion mass, $am\ll 1$, I find
\begin{equation}
E_{V\!\!AC}^0=-\left({7\over 8}\right){\pi^2\over 360}{L^2\over a^3}
+{m\over 24}{L^2\over a^2}+{(eB)^2\over 12\pi^2}L^2a\left[
\ln \left(a\sqrt{eB}\right)+{1\over 8}\right],
\label{E_14}
\end{equation}
notice that this value is twice that obtained in Ref. \cite{Erdas:2010mz}, because in this work I am investigating a Dirac fermion, while Ref. \cite{Erdas:2010mz} investigates a Majorana fermion. Notice also that, instead of the term proportional to $m^2\over a$ that appears in Ref. 
\cite{Erdas:2010mz}, I obtain a term proportional to $2m\over a^2$. It is an interesting physical feature that, for $\beta =0$, the leading mass correction to massless $E_{V\!\!AC}^0$ of an MIT-bag fermion is linear, not quadratic, in $m$. This happens also when the Lorentz violation is spacelike, as we will see later.
 To evaluate the next to leading order term, $E_{V\!\!AC}^1$, since $a/q\gg 1$, I neglect all terms with $n>1$ in the sum of Eq. (\ref{E_13}), and find
\begin{eqnarray}
E_{V\!\!AC}^1&=&\lim_{\epsilon \to 0}{L^2 eB a\over  \sqrt{\pi^3}\Gamma(s)} \left({a\over\sqrt{ {2m\over a} +k^2_1}}\right)^{s-1/2}
\nonumber \\
&\times&\int_0^\infty dz \,z^{s-3/2}e^{-\left(z+z^{-1} \right)\sqrt{ {2m a} +a^2k^2_1}} \coth\left({zaeB\over\sqrt{ {2m\over a} +k^2_1}}\right).
\label{E_16}
\end{eqnarray}
I do the $z$-integration using the saddle point method, take the limit for $\epsilon\rightarrow 0$, and obtain
\begin{equation}
E_{V\!\!AC}^1=-{L^2 eB \over 2\sqrt{\pi^3 a}} \left({2m\over a} +k^2_1\right)^{1/4}
e^{-2\sqrt{ {2m a}} +a^2k^2_1} \coth\left({aeB\over\sqrt{ {2m\over a} +k^2_1}}\right),
\label{E_17}
\end{equation}
where $k_1={2\pi\over q}$.
Notice that $E_{V\!\!AC}^1$ is also exponentially suppressed: the larger the ratio $a/q$, the higher the exponential suppression.
%%%%%%%%%%%%%%%%%%%%%%%%%%%%%%%%%%%%%%%%%%%%%%%%%%%%%%%%%%%%%%%%%%%%%
\subsection{Spacelike Lorentz invariance violation}
\label{2_2}
I start this subsection by examining the case of spacelike Lorentz violation in the $x^3$-direction. I begin by examining the light mass case, and consider first the case $\beta\ne 0$, then will consider the case $\beta=0$. I compare Eq. (\ref{E_pm1}) for $E_\pm$ in the case of timelike Lorentz violation to Eq. (\ref{E_pm2}) for $E_\pm$ in the case of spacelike Lorentz violation in the $x^3$-direction, and notice that when I replace $a$ with $a/(1-\lambda)$ in Eq. (\ref{E_pm1}), I obtain Eq. (\ref{E_pm2}). I also examine the roots of Eq. (\ref{MIT2}) which states the constraint that MIT boundary conditions put on $k_3$ for spacelike Lorentz violation in the $x_3$-direction. According to the Appendix, these roots are
 \begin{equation}
k_n = (n-{1\over 2})\pi+{ma\over 1-\lambda}\left[(n-{1\over 2})\pi
\right]^{-1}+{\cal O}(m^2a^2).
\label{kn2}
\end{equation}
When I compare these values to the roots of Eq. (\ref{MIT}), I see that when I replace $a$ with $a/(1-\lambda)$ in the roots of Eq. (\ref{MIT}), I obtain the roots of Eq. (\ref{MIT2}). This is the same replacement I describe above for $E_\pm$.
Therefore, I make that same replacement in Eq. (\ref{E_10}) for $E_{V\!\!AC}$ in the case of timelike violation and non-vanishing $\beta$, and find $E_{V\!\!AC}$ for the case of spacelike Lorentz violation in the $x^3$-direction and non-vanishing $\beta$, shown below
\begin{eqnarray}
E_{V\!\!AC}&=&-{L^2 eB \over 4\pi^{3/2}} \sqrt{ {1-\lambda\over a}}\left[ 2{(1-\lambda)m \over a}+k^2_0\right]^{1/4}
\nonumber \\
&\times&e^{-2a\sqrt{ 2{(1-\lambda)m \over a}+k^2_0}/(1-\lambda)} \coth\left[{aeB\over (1-\lambda)\sqrt{ 2{(1-\lambda)m \over a}}+k^2_0}\right],
\label{E_18}
\end{eqnarray}
and its massless limit is shown below
\begin{equation}
E_{V\!\!AC}=-{L^2 eB \over 4\pi^{3/2}} \sqrt{ {1-\lambda\over a}}k^{1/2}_0
e^{-2ak_0/(1-\lambda)} \coth\left[{aeB\over (1-\lambda)k_0}\right].
\label{E_18a}
\end{equation}

For $\beta =0$, I make the same replacement in Eq. (\ref{E_14}) and obtain the leading order term, $E_{V\!\!AC}^0$, when $am\ll1$
\begin{equation}
E_{V\!\!AC}^0=-\left({7\over 8}\right){\pi^2\over 360}{L^2(1-\lambda)^3\over a^3}
+{m\over 24}{L^2(1-\lambda)^2\over a^2}+{(eB)^2\over 12\pi^2}{L^2a\over (1-\lambda)}\left[
\ln \left({a\sqrt{eB}\over 1-\lambda}\right)+{1\over 8}\right].
\label{E_19}
\end{equation}
With the same replacement in Eq. (\ref{E_17}), I obtain the next to leading order term
\begin{eqnarray}
E_{V\!\!AC}^1&=&-{L^2 eB \over 2\pi^{3/2}} \sqrt{ {1-\lambda\over a}} \left[2{(1-\lambda)m \over a}+k^2_1\right]^{1/4}
\nonumber \\
&\times&e^{-2a\sqrt{ 2{(1-\lambda)m \over a}+k^2_1}/(1-\lambda)} \coth\left[{aeB\over (1-\lambda)\sqrt{2{(1-\lambda)m \over a}+k^2_1}}\right].
\label{E_21}
\end{eqnarray}

Next, I investigate the case of spacelike Lorentz violation in the $x^1$-$x^2$ direction. I find that, if I compare Eq. (\ref{E_pm1}) to Eq. (\ref{E_pm3}), I need to divide Eq. (\ref{E_pm1}) by $eB$, then replace in it $eB$ with $(1-\lambda/2)^2eB$ and finally multiply it by $eB$ again, and I obtain Eq. (\ref{E_pm3}). Notice that, since $\lambda \ll 1$, I can approximate $(1-\lambda/2)^2\simeq (1-\lambda)$. I make this manipulations in Eq. (\ref{E_10}) for $\beta\ne0$  and small mass, and find $E_{V\!\!AC}$ for spacelike Lorentz violation in the $x^1$-$x^2$ direction,
\begin{equation}
E_{V\!\!AC}=-{L^2 eB \over 4\sqrt{\pi^3 a}} \left(2{m \over a}+k^2_0\right)^{1/4}
e^{-2a\sqrt{ 2{m \over a}}+k^2_0} \coth\left[{a(1-\lambda)eB\over\sqrt{ 2{m \over a}+k^2_0}}\right],
\label{E_24}
\end{equation}
with a massless limit of
\begin{equation}
E_{V\!\!AC}=-{L^2 eB \over 4\sqrt{\pi^3 a}} k^{1/2}_0
e^{-2ak_0} \coth\left[{a(1-\lambda)eB\over k_0}\right].
\label{E_24a}
\end{equation}
For $\beta =0$ and $am\ll1$, I make those replacements in Eq. (\ref{E_14}), and obtain
\begin{equation}
E_{V\!\!AC}^0=-\left({7\over 8}\right){\pi^2\over 360}{L^2\over a^3}
+{m\over 24}{L^2\over a^2}+{(1-\lambda)(eB)^2\over 12\pi^2}L^2a\left[
\ln \left(a\sqrt{(1-\lambda)eB}\right)+{1\over 8}\right].
\label{E_25}
\end{equation} 
The next to leading order term, for spacelike Lorentz violation in the $x^1$-$x^2$ direction, small mass, and $\beta =0$, is obtained from Eq. (\ref{E_17}) by making the replacements described above
 \begin{equation}
E_{V\!\!AC}^1=-{L^2 eB \over 2\sqrt{\pi^3 a}} \left( 2{m \over a}+k^2_1\right)^{1/4}
e^{-2a\sqrt{ 2{m \over a}}+k^2_1} \coth\left[{a(1-\lambda)eB\over\sqrt{ 2{m \over a}+k^2_1}}\right].
\label{E_27}
\end{equation}

Last, I examine spacelike Lorentz violation in the direction of the compactified extra dimension. When  I compare Eq. (\ref{E_pm1}) and Eq. (\ref{E_pm4}), I need to replace $k_\ell$ with $(1-\lambda)k_\ell$ in Eq. (\ref{E_pm1}) to obtain Eq. (\ref{E_pm4}). I make that replacement in Eq. (\ref{E_10}) for $\beta\ne0$  and small mass, and find
\begin{eqnarray}
E_{V\!\!AC}&=&-{L^2 eB \over 4\sqrt{\pi^3 a}} \left[ 2{m \over a}+(1-\lambda)^2k^2_0\right]^{1/4}
e^{-2a\sqrt{ 2{m \over a}+(1-\lambda)^2k^2_0}}\nonumber \\
&\times&
 \coth\left({aeB\over\sqrt{ 2{m \over a}+(1-\lambda)^2k^2_0}}\right),
\label{E_28}
\end{eqnarray}
whose massless limit is
\begin{equation}
E_{V\!\!AC}=-{L^2 eB \over 4\sqrt{\pi^3 a}} \sqrt{(1-\lambda)k_0}\,\,\,
e^{-2a(1-\lambda)k_0}
 \coth\left({aeB\over(1-\lambda)k_0}\right).
\label{E_28a}
\end{equation}
When $\beta =0$, I need to make that replacement in Eq. (\ref{E_14}) to obtain the leading order term, $E_{V\!\!AC}^0$, when $am\ll1$. However, Eq. (\ref{E_14}) does not show dependence on any of the $k_\ell$, so the leading order term for spacelike Lorentz violation in the direction of the compactified dimension is the same as shown in Eq. (\ref{E_14}). The next to leading order term for small mass is found by making the aforementioned replacement in Eq. (\ref{E_17}), to obtain
\begin{eqnarray}
E_{V\!\!AC}^1&=&-{L^2 eB \over 2\sqrt{\pi^3 a}} \left[2{m \over a}+(1-\lambda)^2k^2_1\right]^{1/4}
e^{-2a\sqrt{2{m \over a}+(1-\lambda)^2k^2_1}}\nonumber \\
&\times&
 \coth\left({aeB\over\sqrt{2{m \over a}+(1-\lambda)^2k^2_1}}\right),
\label{E_29}
\end{eqnarray}
Notice how, in Eqs. (\ref{E_28}) - (\ref{E_29}), the Lorentz violation parameter $\lambda$ affects only the part of the Casimir energy dependent on the compactified extra dimension.
%%%%%%%%%%%%%%%%%%%%%%%%%%%%%%%%%%%%%%%%%%%%%%%%%%%%%%%%%%%%%%%%%%%%%
\section{Casimir pressure}
\label{4}
The definition of Casimir pressure is
\begin{equation}
P_C=-{1\over L^2}{\partial E_{V\!\!AC}\over \partial a}.
\label{P_C0}
\end{equation}

When the Lorentz violation is timelike and $\beta\ne 0$, the low mass Casimir pressure is
\begin{equation}
P_C=-{ eB \over 2\sqrt{\pi^3 a^3}} {(m+ak_0^2)\over \left(2{m \over a}+k^2_0\right)^{1/4}}
e^{-2a\sqrt{2{m \over a}+k^2_0}} \coth\left({aeB\over\sqrt{2{m \over a}+k^2_0}}\right),
\label{P_C1}
\end{equation}
and its massless limit is
\begin{equation}
P_C=-{ eB \over 2\sqrt{\pi^3 a}} k_0^{3/2}
e^{-2ak_0} \coth\left({aeB\over k_0}\right),
\label{P_C1a}
\end{equation}
it is attractive and it does not have a dependence on the Lorentz violating parameter. 
For $\beta=0$, I find
\begin{eqnarray}
P_C&=&-\left({7\over 8}\right){\pi^2\over 120a^4}
+
{m\over 12a^3}-{(eB)^2\over 12\pi^2}\left[
\ln \left(a\sqrt{eB}\right)+{9\over 8}\right]- { eB \over \sqrt{\pi^3 a^3}} 
\nonumber \\
&\times& {(m+ak_1^2)\over \left(2{m \over a}+k^2_1\right)^{1/4}}
e^{-2a\sqrt{2{m \over a}+k^2_1}} \coth\left({aeB\over\sqrt{2{m \over a}+k^2_1}}\right),
\label{P_C2}
\end{eqnarray}
valid when $am\ll 1$. 
Dropping terms which are negligible in the small mass approximation, Eq. (\ref{P_C2}) becomes
\begin{equation}
P_C=-\left({7\over 8}\right){\pi^2\over 120a^4}
-{(eB)^2\over 12\pi^2}\left[
\ln \left(a\sqrt{eB}\right)+{9\over 8}\right]-{ eB \over \sqrt{\pi^3 a}} k^{3/2}_1
e^{-2ak_1} \coth\left({aeB\over k_1}\right),
\label{P_C3}
\end{equation}
and it is attractive. 

For spacelike Lorentz violation in the $x^3$-direction, low mass, and $\beta\ne 0$, I find
\begin{eqnarray}
P_C&=&-{ eB (1-\lambda)^{1/2}\over 2\sqrt{\pi^3 a^3}} {[m+ak_0^2/(1-\lambda)]\over \left[2{m(1-\lambda) \over a}+k^2_0\right]^{1/4}}
e^{-2a\sqrt{2{m(1-\lambda) \over a}+k^2_0}/(1-\lambda)} 
\nonumber \\
&\times&\coth\left({aeB\over(1-\lambda)\sqrt{2{m(1-\lambda) \over a}+k^2_0}}\right),
\label{P_C5}
\end{eqnarray}
and its massless limit is
\begin{equation}
P_C=-{ eB \over 2\sqrt{\pi^3 a(1-\lambda)}} k_0^{3/2}
e^{-2ak_0/(1-\lambda)} \coth\left[{aeB\over (1-\lambda) k_0}\right].
\label{P_C5a}
\end{equation}
They display
the same features as the low mass pressure in the case of timelike Lorentz violation and, in addition, a dependence on the Lorentz violating parameter that strengthens the pressure slightly, when compared to the case of timelike violation. 
When $\beta=0$, I find
\begin{eqnarray}
P_C&=&-\left({7\over 8}\right){\pi^2(1-\lambda)^3\over 120a^4}
-{(eB)^2\over 12\pi^2(1-\lambda)}\left[
\ln \left({a\sqrt{eB}\over 1-\lambda}\right)+{9\over 8}\right]-{ eB \over \sqrt{\pi^3 a(1-\lambda)}} k^{3/2}_1
\nonumber \\
&\times&e^{-2ak_1/(1-\lambda)} \coth\left[{aeB\over k_1(1-\lambda)}\right],
\label{P_C6}
\end{eqnarray}
valid when $am\ll 1$ and where I neglected some smaller terms as I did in Eq. (\ref{P_C3}). 

When spacelike Lorentz violation in the $x^1$-$x^2$ direction is present, the pressure for non-vanishing $\beta$ and low mass is
\begin{equation}
P_C=-{ eB \over 2\sqrt{\pi^3 a^3}} {(m+ak_0^2)\over \left(2{m \over a}+k^2_0\right)^{1/4}}
e^{-2a\sqrt{2{m \over a}+k^2_0}} \coth\left[{a(1-\lambda)eB\over\sqrt{2{m \over a}+k^2_0}}\right],
\label{P_C8}
\end{equation}
and its massless limit is
\begin{equation}
P_C=-{ eB \over 2\sqrt{\pi^3 a}} k_0^{3/2}
e^{-2ak_0} \coth\left[{a(1-\lambda)eB\over k_0}\right].
\label{P_C8a}
\end{equation}
The pressure for vanishing $\beta$ is
\begin{eqnarray}
P_C&=&-\left({7\over 8}\right){\pi^2\over 120a^4}
-{(1-\lambda)(eB)^2\over 12\pi^2}\left[
\ln \left(a\sqrt{(1-\lambda)eB}\right)+{9\over 8}\right]-{ eB \over \sqrt{\pi^3 a}} k^{3/2}_1
\nonumber \\
&\times&
e^{-2ak_1} \coth\left[{a(1-\lambda)eB\over k_1}\right],
\label{P_C9}
\end{eqnarray}
when $am\ll1$. Notice also that the violation parameter $\lambda$
weakens the pressure in comparison to the case of timelike Lorentz violation. 

Finally, for Lorentz violation in the direction of the compactified extra dimension and non-vanishing $\beta$, the small mass pressure is
\begin{equation}
P_C=-{ eB \over 2\sqrt{\pi^3 a^3}} {[m+a(1-\lambda)^2k_0^2]\over \left[2{m \over a}+(1-\lambda)^2k^2_0\right]^{1/4}}
e^{-2a\sqrt{2{m \over a}+(1-\lambda)^2k^2_0}} \coth\left({aeB\over\sqrt{2{m \over a}+(1-\lambda)^2k^2_0}}\right),
\label{P_C11}
\end{equation}
with a massless limit of
\begin{equation}
P_C=-{ eB \over 2\sqrt{\pi^3 a}} (1-\lambda)^{3/2}k_0^{3/2}
e^{-2a(1-\lambda)k_0} \coth\left[{aeB\over (1-\lambda)k_0}\right].
\label{P_C11a}
\end{equation}
For $\beta=0$, I find
\begin{eqnarray}
P_C&=&-\left({7\over 8}\right){\pi^2\over 120a^4}
-{(eB)^2\over 12\pi^2}\left[
\ln \left(a\sqrt{eB}\right)+{9\over 8}\right]-{ eB \over \sqrt{\pi^3 a}} \left[(1-\lambda)k_1\right]^{3/2}
\nonumber \\
&\times&
e^{-2a(1-\lambda)k_1} \coth\left[{aeB\over (1-\lambda)k_1}\right],
\label{P_C12}
\end{eqnarray}
valid when $am\ll 1$. 
%%%%%%%%%%%%%%%%%%%%%%%%%%%%%%%%%%%%%%%%%%%%%%%%%%%%%
\section{Discussion and conclusions}
\label{5}
In this work I use the zeta function technique to study the Casimir effect due to a lightly massive and charged fermion field that violates Lorentz symmetry in a CPT-even aether-like manner. This quantum field "lives" in a four-dimensional space-time plus a toroidally compactified fifth dimension. A uniform magnetic field is present and is perpendicular to the Casimir plates. The quantum field satisfies MIT bag boundary conditions on the two plates. These boundary conditions impose that the momentum component perpendicular to the plates be discretized through a transcendental equation, and in the Appendix I find all analytic solutions to the transcendental equation for the cases of light, $ma\ll 1$, fermion mass.

Under the assumption that the size of the compactified extra dimension is smaller than the plates distance, $q\ll a$, I obtain simple and accurate analytic expressions of the Casimir energy, $E_{V\!\!AC}$, for non vanishing values of the compactification parameter $\beta$  and for light, and vanishing fermion mass, in the case where the Lorentz violation is timelike, Eqs. (\ref{E_10}), (\ref{E_10a}), when the Lorentz violation is spacelike and parallel to the magnetic field, Eqs. (\ref{E_18}), (\ref{E_18a}), when it is spacelike and perpendicular to the magnetic field, Eqs. (\ref{E_24}), (\ref{E_24a}), and when it is in the direction of the compactified extra dimension, Eqs. (\ref{E_28}), (\ref{E_28a}). For each of the cases listed above, I also obtained the Casimir pressure in Sec. \ref{4}, the one quantity measurable in the lab. These Casimir pressure results are also simple and accurate analytic expressions.
The case of the vanishing compactification parameter needed to be examined separately. In this case, for timelike Lorentz violation, I obtained the Casimir energy for light fermion mass, $am\ll 1$, in Eqs. (\ref{E_14}) for the leading order term, Eq. (\ref{E_17}) for the next to leading order term. 
Next I pursued the cases of spacelike Lorentz violation and calculated the Casimir energy for Lorentz violation parallel to the magnetic field for light mass, Eqs. (\ref{E_19}) and (\ref{E_21}). Then, I went on to obtain the Casimir energy for Lorentz violation perpendicular to the magnetic field for light fermion mass, Eqs. (\ref{E_25}) and (\ref{E_27}). Finally, I calculated the Casimir energy for Lorentz violation in the direction of the compactified extra dimension and light mass, Eqs. (\ref{E_14}) and (\ref{E_29}).

I find that, in the case of timelike Lorentz violation and spacelike Lorentz violation in all directions examined the Casimir pressure is attractive for both cases of light mass and zero mass. It is also interesting that, in the case of Lorentz violation in the direction of the compactified extra dimension, the dependence on the Lorentz violating parameter, $\lambda$, appears only in the part of the Casimir energy and pressure that depends on the size of the compactified extra dimension.
%%%%%%%%%%%%%%%%%%%%%%%%%%%%%%%%%%%%%%%%%%%%%%%%%%%%%
\section{Appendix}
\label{6}
In this Appendix I evaluate analytically the roots of Eqs. (\ref{MIT}) and (\ref{MIT2}) for the case of $ma\ll 1$. I use the notation $z = k_3a$ and $\mu = ma$ and rewrite Eq. (\ref{MIT}) as
\begin{equation}
\mu\sin (z)+z\cos(z) =0.
\label{A1}
\end{equation}
When $\mu=0$
 the roots of Eq. (\ref{A1}) are
\begin{equation}
z_n=(n-{1\over 2})\pi,
\label{A2}
\end{equation}
 with $n\ge 1$. When $\mu\ll 1$ the roots of Eq. (\ref{A1}) are
 \begin{equation}
{\tilde z}_n=z_n+x_n,
\label{A3}
\end{equation}
where $x_n\ll 1$. To find $x_n$, I replace $z$ with ${\tilde z}_n$ in Eq. (\ref{A1}) and find
\begin{equation}
\mu\cos (x_n)-{\tilde z}_n\sin(x_n) =0,
\label{A4}
\end{equation}
solve for $x_n$ and obtain
\begin{equation}
\tan (x_n)={\mu\over {\tilde z}_n}.
\label{A5}
\end{equation}
Since both $\mu\ll 1$ and $x_n\ll 1$, I rewrite the equation above as
\begin{equation}
x_n\simeq{\mu\over z_n}+{\cal O}(\mu^2),
\label{A6}
\end{equation}
and so the roots of Eq. (\ref{A1}) when $\mu\ll 1$, are
\begin{equation}
{\tilde z}_n=z_n+{\mu\over z_n}+{\cal O}(\mu^2).
\label{A7}
\end{equation}
Eq. (\ref{MIT2}), for the case of spacelike violation in the $x^3$ direction, is very similar to Eq. (\ref{MIT}) and, using the same notation, can be written as
\begin{equation}
\mu\sin (z)+(1-\lambda)z\cos(z) =0.
\label{A8}
\end{equation}
Using the same method I use above, the roots of Eq. (\ref{A8}) when $\mu\ll 1$, are found to be
\begin{equation}
{\tilde z}_n=z_n+(1-\lambda)^{-1}{\mu\over z_n}+{\cal O}(\mu^2).
\label{A9}
\end{equation}
%%%%%%%%%%%%%%%%%%%%%%%%%%%%%%%%%%%%%%%%%%%%%%%%%%%%%

%%%%%%%%%%%%%%%%%%%%%%%%%%%%%%%%%%%%%%%%%%%%%%%%%%%%%%%%%%%%%%%%%%%
\end{document}